\documentstyle[12pt]{article}

\def\ai{\'{\i}}
\def\itt{\int_{\tau_1} ^{\tau_2}}
\def\qb{\overline Q}
\def\pp{\pi_\phi}

\def\po{\pi_\Omega}
\def\om{\Omega}
\def\33{e^{3\Omega}}
\def\66{e^{6\Omega}}

\def\e-3{e^{-3\Omega}}

\def\op{e^{6(\Omega+\phi)}}

\def\pb{\overline P}
\begin{document}
\baselineskip.3in

\centerline{\large{\bf Quantization of minisuperspaces as ordinary gauge systems}}

\bigskip

\centerline{  Claudio Simeone\footnote{{\bf e-mail:} simeone@tandar.cnea.edu.ar}}

\medskip

\centerline{\it Departamento de F\ai sica, Comisi\'on Nacional de Energ\ai a At\'omica}

\centerline{\it Av. del Libertador 8250, 1429 Buenos Aires, Argentina}

\centerline{\it and}

\centerline{\it Departamento de F\ai sica, Facultad de Ciencias Exactas y Naturales}

\centerline{\it Universidad de Buenos Aires,  Ciudad Universitaria, Pabell\'on I}

\centerline{\it 1428, Buenos Aires, Argentina.}

\vskip1cm

ABSTRACT

\bigskip
 
Simple cosmological models are used to show that gravitation can be quantized as an ordinary gauge system if the  Hamilton-Jacobi equation for the model under consideration is separable. In this situation, a canonical transformation can be performed such that in terms of the new variables the model has a linear and homogeneous constraint, and therefore canonical gauges are admissible in the path integral. This has the additional practical advantage that gauge conditions that do not generate Gribov copies are then easy to choose.

\vskip1cm


\newpage

I. INTRODUCTION
\bigskip

The evolution of a cosmological model is given by the dynamics of the  fields $g_{\mu\nu}(X)$ and of the matter field $\phi$, which yield from the extremal condition imposed on the Einstein action. The change from the variables $g_{\mu\nu}$ to the set

$$g_{ab}({\bf x}, \tau),\ \ \  N({\bf x}, \tau),\ \ \  N^a({\bf x} , \tau)$$
where $N$ and $N^a$ are the lapse and shift functions defined by Kucha\v r [1]
as a generalization of those introduced by Arnowitt, Deser and Misner [2] $N=(-g^{00})^{-1/2},\  N^a=g^{ab}g_{b0}$, gives the Einstein action the form [3]

$$S[g_{ab}, p^{ab}, N, N^a, \phi]=\int d\tau\int d^3x\,(p^{ab}\dot g_{ab}-N{\cal H} -N^a{\cal H}_a)\eqno(1)$$
where the hamiltonian and momentum constraints
$${\cal H}={1\over 2}G_{abcd}\,p^{ab}p^{cd}-( ^3g)^{1/2}( ^3R-2\Lambda)+{\cal H}_{matt}\approx 0$$
$${\cal H}_a=-2\,g_{ac}\nabla_d\,p^{cd}+{\cal H}_{a,matt}\approx 0\eqno(2)$$
reflect the general covariance of the theory.

The restriction to minisuperspace models and the choice of an homogeneous lapse and zero shift lead to a single constraint of the form [4]

$$H={1\over 2}f_{ij}\,p^ip^j+V(q)\approx 0. \eqno(3)$$
where $f_{ij}$ is the reduced version of the DeWitt supermetric. The constraint $H$ is a generator of gauge transformations giving rise to an infinite number of physically equivalent paths in phase space. If the action

$$S[q^i,p_i,N]=\int d\tau\,(p_i\dot q^i -NH)\eqno(4)$$
is used to quantize the system by means of a path integral, the propagator diverges because of the sum over equivalent paths. This is solved by imposing a gauge condition that selects one path from each class of equivalent paths. Admissible gauge conditions are those which can be reached from any path by performing a gauge transformation that does not modify the action, so that a first restriction on admissible gauge conditions appears. A second -and general- condition is that gauge fixing must not generate Gribov copies, that is, the orbits (i.e. the points of the phase space connected by gauge transformations) must have no more than one intersection with the surface defined by the gauge.

Under the gauge transformation
$$\delta_\epsilon q^i=\epsilon (\tau )[q^i,H], \ \ \  \delta_\epsilon p_i=\epsilon (\tau )[p_i,H], \ \ \  \delta_\epsilon N=\dot\epsilon (\tau), \eqno(5)$$
the action changes by

$$\delta_\epsilon S=\left[ \epsilon (\tau )\left( p_i{\partial H\over\partial p_i}-H\right) \right]_{\tau_1} ^{\tau_2} \eqno(6)$$
so that for a quadratic hamiltonian like that of equation (4) $\delta_\epsilon S$ vanishes only if
$$\epsilon (\tau_1 )=0=\epsilon(\tau_2 ). \eqno(6)$$
This introduces  a restriction on admissible gauges: derivative gauge conditions like
$$\chi\equiv \dot N-\chi_0(q^i,p_i,N)=0 \eqno(7)$$
should be used [5] [6], reflecting a difference between gravitation and ordinary gauge systems, which have linear and homogeneous constraints so that $\delta_\epsilon S$ vanishes even if the boundaries are not mapped onto themselves, allowing then for canonical gauge conditions 
$$\chi (q^i,p_i,\tau )=0. \eqno(8)$$ 
However, a system with a quadratic hamiltonian can be made gauge invariant at the endpoints improving the action with endpoint terms [7]. In a previous paper [8] we have shown that these terms  can be seen  as the result of a canonical transformation which identifies the constraint with one of the new momenta, yielding an action like that of an ordinary gauge system. In the present work we apply this idea to some minisuperspace models; we solve the Hamilton-Jacobi equation for them, and find the gauge-invariant action which makes simple canonical gauges admissible in the path integral. We show that this gauge fixing procedure makes it easy to avoid the Gribov problem, and, from a different point of view, we find that the path integral resulting after the gauge choice clearly manifests the -arbitrary- separation between true degrees of freedom and time.

\vskip1cm

II. GAUGE-INVARIANT ACTION

\bigskip

 Let us consider the canonical transformation $(q^i, p_i)\to (\qb^i,\pb_i)$ generated by the solution $W(q^i,\alpha_\mu ,E)$ of the $\tau$-independent Hamilton-Jacobi equation

$$H\left( q^i,{\partial W\over\partial q^i}\right) =E\eqno(8)$$
matching the constants  $(\alpha_\mu ,E)$ to the new momenta $(\pb_\mu ,\pb_0).$ The new coordinates and momenta verify 

$$[\qb^{\mu},\pb_0]=[\qb^{\mu},H]=0$$

$$[\pb_{\mu},\pb_0]=[\pb_{\mu},H]=0$$

$$[\qb^{0},\pb_0]=[\qb^{0},H]=1.\eqno(11)$$
$\qb^{\mu}$ and $\pb_\mu$ do not change under a gauge transformation and are then called observables, while $\qb^0$ has non-zero bracket with the constraint, suggesting its use to fix the gauge: in terms of the new variables the constraint is linear and homogeneous in the momenta, so that canonical gauges are admissible, and the condition that guarantees that the surface $\chi=0$ is not tangent to the orbits [9] (what will be used to ensure that it does not intersect them more than once) is fulfilled by $\chi\equiv\qb^0$:

$${\textstyle det}([\chi ,H])=[\qb^{0},H]\not=0.\eqno(12)$$

A second transformation, now in the space of observables, generated by 
$$F=P_0\qb^0+f(\qb^\mu ,P_\mu ,\tau) \eqno(13)$$
yields a new non-zero hamiltonian
$$K=NP_0+{\partial f\over\partial\tau}\eqno(14)$$
and a new set of non-conserved observables $(Q^\mu,P_\mu)$, that are therefore appropriate to characterize the trajectories in phase space. As a functional of $Q^i$ and $P_i$ the new gauge-invariant action reads
$${\cal S}=\itt \left( P_i{dQ^i\over d\tau}-NP_0-{\partial f\over\partial\tau}\right) d\tau\eqno(15)$$
so that the system is now an ordinary gauge system, i.e. one with a linear an homogeneous constraint $P_0\approx 0$ and a non-zero hamiltonian ${\partial f\over\partial\tau}$. The path integral has then the form 
$$\int DQ^0\, DP_0\, DQ^\mu\, DP_\mu\, DN\, \delta(\chi )\,\vert [\chi,P_0]\vert\,  e^{i \itt \left( P_i{dQ^i\over d\tau}-NP_0-{\partial f\over\partial\tau}\right) d\tau}\eqno(16)$$
where $\vert [\chi, P_0]\vert$ is the Fadeev-Popov determinant, and admits any canonical gauge. In terms of the original variables the gauge invariant action is writen
$${\cal S}=\itt\left( p_i{dq^i\over d\tau }-NH\right) d\tau +B\eqno(17)$$
with [8]
$$B=\left[ \qb^i\pb_i -W+Q^\mu P_\mu-f\right]_{\tau_1}^{\tau_2}.\eqno(18)$$
The existence of a gauge such that 
$\tau =\tau (q^i)$ (intrinsic time) [10] and the vanishing of  the endpoint terms $B$  in that gauge and on the constraint surface  assure that the new  action $\cal S$ and the original one $S$ weigh the paths in the same way [8]. This requirement determines the function $f$ in equation (13).

\vskip1cm

III. MINISUPERSPACES

\bigskip

The procedure we have shown depends on the possibility to find a solution of the Hamilton-Jacobi equation for the system. The  separability of this equation and its application to cosmological models has been widely studied [11] [12]. Here we will study simple models,  for which the Hamilton-Jacobi equation can be easily solved.
Consider an isotropic and homogeneous Friedmann-Robertson-Walker (FRW) metric

$$ds^2=N^2d\tau^2-a^2(\tau)\left({dr^2\over 1-kr^2}+r^2d\theta^2+r^2sin^2\theta d\varphi^2\right).$$
The hamiltonian constraint 

$$H={1\over 4}\e-3 (\pp^2-\po^2)+\Lambda\33 \approx 0,\eqno(19)$$
with $\om\sim \ln a(\tau)$ corresponds to a FRW metric with massless scalar field, non-zero cosmological constant and null curvature. The evolution is restricted to one of the two surfaces
$$\po=\pm\sqrt{\pp^2+4\Lambda\66}$$
separated by $\po=0$; from a geometrical point of view this means that the topology of the surface where the gauge choice must select only one point from each orbit is that of half a plane. The  $\tau$-independent Hamilton-Jacobi equation has the solutions

 $$W_\pm =\pb\phi\pm\int d\om\sqrt{\pb^2-4\pb_0 \33 + 4\Lambda\66}\eqno(20)$$
which are of the form
$$W=w(q^0,\pb_0,\pb)+C(q^i)\pb\eqno(21)$$
with $q^0=\om ,\  q=\phi$. Fixing the gauge by means of the canonical condition

$$\chi\equiv\qb^0-g(\pb ,T(\tau ))=0\eqno(22)$$
with $T(\tau)$ a monotonous function of $\tau$, as  $\qb^0 ={\partial W_\pm\over\partial \pb_0}=\qb^0(q^0,\pb_0,\pb),$ if we choose 

$$g(\pb ,T(\tau ))=\qb^0(q^0=T(\tau ),\pb_0=0,\pb )$$ 
in terms of the original variables we have 
$$q^0=T(\tau ).\eqno(23)$$
The surface $\chi=0$ is thus a plane $\om=constant$ for each value of $\tau$. This guarantees that the gauge (22) does not produce Gribov copies, because if it did so at any $\tau$, at another one it would be $[\chi, H]=0$, which is prevented by the gauge fixing procedure.

The additional endpoint terms $B$ for the action of this system vanish in the gauge (22) if 
$$f=\qb P-w(T(\tau),\pb_0=0,\pb=P).$$
Thus, in this  gauge and on the constraint surface we have
$$Q={\partial f\over\partial P}=C(q^i)$$  
which, together with equation (23), means that $Q$ and $\tau$ define a hypersurface in the original configuration space. The hamiltonian for the reduced system $(Q,P)$ is 
$${\partial f\over\partial\tau}=\mp\sqrt{P^2+4\Lambda e^{6T}}\,{dT\over d\tau},$$
and it resembles that of a relativistic particle of $T$-dependent mass $m=2\Lambda^{1/2}e^{3T}$. In the gauge (22) and on the constraint surface the path integral for the system has then the simple form 

$$<\phi'',\om''\vert\phi',\om' >=\int DQDP\, e^{i\int_{T'} ^{T''} \left(PdQ\pm\sqrt{P^2+4\Lambda e^{6T}}dT\right)}, \eqno(24)$$
where the endpoints are given by $T'=\om'$ and $T''=\om''$, while the paths go from $Q'=\phi'$ to $Q''=\phi''$.

\bigskip

Another easily solvable constraint is 

$$H ={1\over 4}\e-3 (\pp^2-\po^2)+\33 V(\phi)\approx 0\eqno(25)$$
with $V(\phi)=\lambda e^{6\phi}$, which corresponds to a FRW metric with a massless scalar field and an exponential potential which can reproduce an increasing cosmological constant. The topology is again like that of two disjoint planes. Multiplying by the positive-definite function $\33$ an equivalent constraint is obtained:
$$ H'={1\over 4}(\pp^2-\po^2)+\lambda\op \approx 0\eqno(26)$$
This is an example in which the new momentum $\pb_0$ differs from the constraint $H$ by a non-zero factor, so that the bracket of $\qb$ with the constraint is zero only on the constraint surface, but $\qb$ is still an observable.
The solution of the $\tau$-independent Hamilton-Jacobi equation is 

$$W=\pb(\phi-\om ) +{1\over\pb}\left(\pb_0(\phi+\om )-{\lambda\over 6}\op \right) \eqno(27)$$
and has the form (21) if $q^0=\phi+\om$ and $q=\phi-\om$. The canonical gauge

$$\chi\equiv\qb^0-{T(\tau )\over\pb}=0\eqno(28)$$ 
gives
$$q^0=\phi+\om=T(\tau),$$
corresponding to a plane $\phi+\om= constant$ for each value of $\tau$. This gauge does not generate Gribov copies, because if at any $\tau$ the plane intersected an orbit, at another $\tau$ they should be tangent,  yielding $[\chi,H]=0$. As in the former example, with the choice of the function $f$ such that the endpoints $B$ vanish in the gauge (28) the new coordinate $Q$ and $\tau$ define a hypersurface in the original configuration space. The hamiltonian for the reduced system described by $Q$ and $P$ is

$${\partial f\over\partial\tau}={\lambda\over P}e^{6T(\tau )}{d T\over d\tau},$$
and the path integral in the gauge (28) and on the constraint surface has the form

$$<\phi'',\om''\vert\phi',\om' >=\int DQDP\,e^{i\int_{T'} ^{T''} 
\left(PdQ-{\lambda\over P}e^{6T}dT\right) },\eqno(29)$$
where $T'=\phi'+\om'$ and $T''=\phi''+\om''$, and the paths go from $Q'=\phi'-\om'$ to $Q''=\phi''-\om''$. As before, the evolution  of the system described by the observables $(Q,P)$ is given by a $T$-dependent hamiltonian.

\bigskip

In a more general case [8], whenever the hamiltonian constraint has the form
$$H=G(\phi,\om)(\pp^2-\po^2)+V(\phi,\om)\approx 0  \eqno(30)$$
with $G$ and $V$ positive definite functions such that ${V(\phi,\om)\over G(\phi,\om)}=L_1(\phi+\om) L_2(\phi-\om)$ the change to the null coordinates
$$u=R_1(\phi-\om),\ \  v=R_2(\phi+\om)$$
with the choice $L_2(z)=R'(z)$ yields the equivalent constraint
$$H'=\pi_u\pi_v+{1\over 4}L(v)\eqno(31)$$
related to $H$ by $H=V(\phi,\om)H'$, with $L(v)\equiv{L_1(v)\over L_2(v)}$. The generator function $W$ which identifies the set $(\qb^i,\pb_i)$ is
$$W=\pb u+{\pb_0\over\pb}v-{1\over 4\pb}\int L(v)dv,\eqno(32)$$
and has the form
$$W=w(v,\pb_0,\pb)+C(u)\pb.\eqno(33)$$
The canonical gauge
$$\chi\equiv\qb^0-{T(\tau )\over\pb}=0\eqno(34)$$ 
is equivalent to
$$v=R_2(\phi+\om)=T(\tau).$$
Gribov problem is avoided in the same way it was done before, as the constraint hypersurface splits in 
$$\po=\pm\sqrt{{V(\phi,\om)\over G(\phi,\om)}+\pp^2}.$$
The generator function $f$ giving the change to nonconserved observables is
$$f=\qb P+{1\over 4P}\int L(T) dT\eqno(35)$$
and the new variables are related to the null ones by
$$Q^0={v\over P}\ \ \ \ \ \ Q=u+{v\over P^2}(1-P_0)-{1\over 4P^2}\int L(T)dT$$
$$\pi_u=P\ \ \ \ \ \ \pi_v={P_0-L(v)\over P}.\eqno(36)$$
The system $(Q,P)$ is governed by the non-zero hamiltonian ${\partial f\over\partial\tau}={1\over 4P} L(T){dT\over d\tau},$ so that the path integral on the constraint surface and in the gauge (34) is
$$<\phi'',\om''\vert\phi',\om' >=\int DQDP\,e^{i\int_{T'} ^{T''} 
\left(PdQ-{L(T)\over 4P}dT\right) },\eqno(37)$$
where $T'=R_2(\phi'+\om')$ and $T''=R_2(\phi''+\om'')$, and the paths go from 
$Q'=R_1(\phi'-\om')$ to $Q''=R_1(\phi''-\om'')$. The evolution of the reduced system is  given by a $T$-dependent hamiltonian.  As in the preceeding examples, this is nothing more, however, than a consequence of the  fact that the gauge choice is not only a procedure to remove divergences from the path integral, but also a reduction procedure to physical degrees of freedom, giving rise to time and dynamical evolution [3]. This can  soon be realized by examining the result of a ``wrong'' gauge choice, like a $\tau$-independent one.  We would  then obtain a zero hamiltonian for the reduced system, yielding no dynamical evolution.

\vskip1cm

IV. CONCLUSIONS

\bigskip

The presence of a quadratic constraint in the action of gravitation would make impossible, in principle, the quantization of cosmological models by means of the usual path integral procedure for gauge systems [6]. However, when the Hamilton-Jacobi equation for the system can be solved, this can mean nothing more than an unappropriate choice of coordinates and momenta. The change to variables such that the constraint is linear and homogeneous in the momenta provides the action with invariance at the endpoints, making canonical gauges admissible in the path integral. In terms of the original variables the gauge invariant action has additional endpoint terms, but these terms do not modify the dynamical trajectories of the system. The vanishing of the endpoint terms  in a gauge that  defines a hypersurface in the original configuration space guarantees that the improved action weighs the paths in the same way as the original one. However, because the new action is gauge invariant,  the path integral can be computed in any canonical gauge. Canonical gauges make it easy  to avoid Gribov copies, and also to define a global time [3].

It should be emphasized that our procedure  will only work in a limited class of minisuperspace models; even in the case of homogeneous and isotropic models, for an arbitrary potential $V(\phi,\Omega)$ -yielding a topologically nontrivial constraint surface- it is most unlikely to work in general. However, the Hamilton-Jacobi equation is separable for constraints more general than those considered here: for example, solutions are known for models with more than one scalar field, or with a potential $V(\phi ) \sim{1\over \phi^2},$ which reproduces an inflationary model with a decaying cosmological constant [12].
The existence of an intrinsic time and the possibility to avoid the Gribov problem in these and other separable models should be the object of a further analysis.

\vskip1cm

ACKNOWLEDGEMENTS

\bigskip

I wish to thank Rafael Ferraro for reading  the manuscript and making helpful comments.

\vskip1cm

REFERENCES

\bigskip

\noindent 1. K. V. Kucha\v r,  J. Math. Phys.  {\bf 17}, 777 (1976).

\medskip

\noindent 2. R. Arnowitt, S. Deser and C. Misner, in {\it Gravitation, an Introduction to Current Research}, ed. L. Witten, Wiley, New York (1962).

\medskip

\noindent 3. A. O. Barvinsky, Phys. Rep. {\bf 230}, 237 (1993).

\medskip

\noindent 4. J. J. Halliwell, in {\it Introductory Lectures on Quantum Cosmology, Proceedings of the Jerusalem Winter School on Quantum Cosmology and Baby Universes}, ed. T. Piran, World Scientific, Singapore (1990). 

\medskip
 
\noindent 5. C. Teitelboim, Phys. Rev. D {\bf 25}, 3159 (1982).

\medskip
 
\noindent 6. J. J. Halliwell, Phys. Rev. D {\bf 38}, 2468 (1988).

\medskip

\noindent 7. M. Henneaux, C. Teitelboim  and J. D. Vergara, Nucl. Phys. B {\bf 387}, 391 (1992).

\medskip

\noindent 8. R. Ferraro  and C. Simeone, J. Math. Phys, {\bf 38}, 599 (1997). 

\medskip

\noindent 9. M. Henneaux  and C. Teitelboim, {\it Quantization of Gauge Systems}, Princeton University Press, New Jersey (1992).

\medskip

\noindent 10. K. V. Kucha\v r, in {\it Proceedings of the 4th Canadian Conference on General Relativity and Relativistic Astrophysics}, eds. G. Kunstatter, D. Vincent and J. Williams, World Scientific, Singapore  (1992).

\medskip  

\noindent 11. D. S. Salopek and J. R. Bond, Phys. Rev. D {\bf 42}, 3936 (1990).

\medskip

\noindent 12. D. S. Salopek and J. M. Stewart, Class. Quantum Grav. {\bf 9}, 1943 (1992).

\medskip

\end{document}